\documentclass[aps,pre,onecolumn,epsf,graphicx]{revtex4}

\def\bec{\begin{center}}
\def\enc{\end{center}}

\def\ben{\begin{equation}}
\def\ba{\begin{array}}
\def\bea{\begin{eqnarray}}

\def\een{\end{equation}}
\def\eea{\end{eqnarray}}
\def\ea{\end{array}}
\def\btab{\begin{table}}
\def\btabu{\begin{tabular}}
\def\etab{\end{table}}
\def\etabu{\end{tabular}}
\def\bit{\begin{itemize}}
\def\eit{\end{itemize}}
\def\bef{\begin{figure}[htb]}
\def\befh{\begin{figure}[!h!]}
\def\enf{\end{figure}}

\def\b1{{\bf 1}}

\def\cos{\hbox{cos}\:}

\usepackage{epsfig,latexsym}

\newcommand \bew {\begin{widetext}}
\newcommand \enw {\end{widetext}}

\begin{document}

\title{\bf\noindent The two loop calculation of the disjoining
pressure of a symmetric electrolyte soap film}

\author{D.S. Dean$^{(1,2)}$ and R.R. Horgan$^{(1)}$}

\affiliation{
(1) DAMTP, CMS, University of Cambridge, Cambridge, CB3 0WA, UK \\
(2) IRSAMC, Laboratoire de Physique Quantique, Universit\'e Paul Sabatier, 118 route de Narbonne, 31062 Toulouse Cedex 04, France\\
E-Mail:dean@irsamc.ups-tlse.fr, rrh@damtp.cam.ac.uk}

\date{25 February 2004}
\begin{abstract}
In this paper we consider the two-loop calculation of the disjoining
pressure of a symmetric electrolytic soap film. We show that the disjoining 
pressure is finite when the loop expansion is resummed using a cumulant 
expansion and requires no short distance cut-off. The loop expansion
is resummed in terms of an expansion in $g= l_B/l_D$ where $l_D$ is
the Debye length and $l_B$ is the Bjerrum length. We show that there
there is a non-analytic contribution of order $g\ln(g)$. We also show that the 
two-loop correction is greater than the one-loop term at large film thicknesses 
suggesting a non-perturbative correction to the one-loop result in this limit. 
\end{abstract}  
\maketitle
\vspace{.2cm}
\pagenumbering{arabic}
\section{Introduction}
The determination of the effective interaction between two surfaces of a film-like structure
is essential for the understanding of the conformational stability
of a whole range of systems in soft-condensed matter physics. Examples
are found in interactions between membranes and colloid physics
\cite{books}. The effective interaction between  two semi-infinite 
dielectrics separated by a vacuum was calculated by Lifshitz \cite{lif} and
the effect of an intervening dielectric was determined by Dzyaloshinski,
Lifshitz and Pitaevskii \cite{dlp} and subsequently reformulated more 
simply in \cite{van}. The effective interaction is due to
van der Waals or dispersion forces and is given by the sum over
the Matsubara modes of the problem; the non-zero frequency modes are a 
quantum effect and the contribution of zero modes corresponds to a thermal Casimir type 
effect. In purely dielectric problems the contribution of each mode
is described by a free Gaussian field theory and it has been shown how
the input to these free field theories can be determined using dielectric
data, see \cite{mani} and references therein.  

In many situations, especially in biology, the middle slab of the system
is filled with an electrolyte. For example, this is the case
for a simple soap film connected to a bulk which is filled with an aqueous salt solution. 
It has been argued \cite{dani} that the presence of an electrolyte will
not effect the non-zero frequency contribution to the interaction basically 
because the response time of the ions is too large to couple them to the these
non-zero frequency modes. However the zero-frequency, or static, component
of the interaction does couple to the ionic distribution and consequently its contribution
is strongly affected by the presence of electrolyte. The field theory describing the 
zero frequency fluctuations of the electric field in the presence of ions is no longer a free
field theory; the interactions are made up of two components: the basic 
thermal fluctuations of the electrostatic field and the induced Coulombic
interactions between the ions. In the presence of a neutral surface and
a symmetric monovalent electrolyte the mean-field electrostatic potential
is zero and the one-loop contribution about this trivial mean-field solution
is equivalent to Debye-H\"uckel theory. The one-loop result is described and
computed in \cite{mani} and has been rederived using a variety of 
different methods, for instance see \cite{netz1,deho}. In the absence 
of electrolyte the zero frequency contribution to the effective interaction, commonly known as the
disjoining pressure $P_d$, behaves as $P_d \approx -A/l^3$ for a film of thickness
$l$. In the presence of electrolyte, at large film separations relative to the 
Debye length $l_D$, the disjoining pressure is screened at one loop
and has the form  $P_d \approx -A\exp(-2l/l_D)/l$. In both of these cases
the constant $A$ is positive and hence the effective force between the surfaces is attractive 
at one-loop order. 

In this paper we calculate $P_d$ to two-loop order. The naive two-loop
calculation is in fact divergent when the dielectric constant of the film, or slab, is greater 
than that of the exterior \cite{netz2}, but we show here that the result
can be resummed using a cumulant expansion similar to that used in \cite{cum} for the bulk 
to give a finite result which we argue is to be expected on physical grounds. The expansion we use 
can be controlled systematically, and the perturbative parameter is $g= l_D/l_B$ where
$l_B$ is the Bjerrum length. While the one-loop or first term in the
expansion in $g$ is regular, the second order term in $g$ is of the form
$g^2\ln(g)$, this singular behavior is found in the Onsager-Samaras
limiting law for the excess surface tension of electrolyte solutions and
has its origin in the image forces which repel the ions from the air/water
interface. We also analyse the behavior of the $O(g^2)$ term for large
film thickness $l$ and find that this term has the behavior
$-A'\ln(l/l_D)\exp(-2l/l_D)/l$, which means that it becomes larger than the
one-loop result for very thick films. We conjecture that this result suggests that
it is the first term in a series which sums to give a contribution at large $l$ to the
disjoining pressure with behaviour $P_d \approx -A''\exp(-2l/l_D)/l^\alpha$,
where the exponent $\alpha$ has the form $\alpha = 1 + \alpha_1 g + \alpha_2 g^2 \cdots$. 
    
\section{Two-loop calculation}
The model of the soap film we shall consider is the following. The 
film body is a planar slab of thickness $l$ surrounded on both sides
by an external dielectric medium such as air. The interior of the
film is filled with electrolyte and connected to a bulk reservoir 
of electrolyte. The electrolyte consists of a passive solvent of uniform
dielectric constant containing monovalent anions and cations 
which are treated as point charges of  charges $-e$ and $e$ respectively. 
The electrolyte is symmetric in the sense that the anions and cations are 
identical in every respect other than that they have opposite electric charges.

In what follows we shall use the standard field theoretic formulation
for symmetric monovalent electrolyte systems \cite{ft}.
The grand partition function for a film of electrolyte solution, surrounded
on both sides by an external medium such as air, of thickness $l$ is given by 
\begin{equation}
\Xi_F(l) = \int d[\phi] \exp\left( S_F[\phi,l]\right)\;,
\end{equation}
where the action $S_F$ of the film is given by
\begin{equation}
S_F[\phi,l] = 
-{\beta \epsilon\over 2}\int_{F_I} d{\bf x} \ \left(\nabla\phi\right)^2
+ 2 \mu \int_{F_I} d{\bf x}\  \cos\left(e\beta\phi\right) -
{\beta \epsilon_0\over 2}\int_{F_E} d{\bf x} \ \left(\nabla\phi\right)^2.
\end{equation}
The region inside the film is denotes $F_I$ and, if $z$ is the coordinate perpendicular 
to the film surface, then $F_I$ is the region $z \in [0,l]$. Inside $F_I$ the dielectric 
constant is $\epsilon$ and is taken to be that of the solvent, in most cases water, 
and the fugacities for the cations and anions in this symmetric
electrolyte are the same and are denoted by $\mu$. The region outside the film is denoted
by $F_E$; in $F_E$ there is no electrolyte and the dielectric constant is
denoted by $\epsilon_0$. In this paper we consider the physically relevant case where
$\epsilon > \epsilon_0$.

For a bulk solution  of thickness $l$ the grand partition function is given by
\begin{equation}
\Xi_B(l) = \int d[\phi] \exp\left( S_B[\phi,l]\right)\;,
\end{equation}
where $S_B$ is the bulk action 
\begin{equation}
S_B = -{\beta \epsilon\over 2}\int_B d{\bf x} \ \left(\nabla\phi\right)^2
+ 2 \mu \int_B d{\bf x}\  \cos\left(e\beta\phi\right)\;. 
\end{equation}
Here there is no contact with an exterior region and one can close the system
by using periodic boundary conditions in the $z$ direction. 
The value of the fugacity is determined by the bulk density $\rho$ and is given by
\begin{equation}
\mu = {\rho\over \langle \cos(e\beta\phi)\rangle_B} = Z\rho\;. \label{zed}
\end{equation}
Here, $Z^{-1} =  \langle \cos(e\beta\phi)\rangle_B$ is a renormalization factor accounting
for ion self-interactions, and the expectation is taken at an arbitrary position in an infinite
bulk system, {\em i.e.} where $l\to \infty$. In experiments on planar systems or films 
the physical quantity which is measured is the disjoining pressure which is due to the 
effective interaction induced by the presence of the film surfaces. The disjoining pressure 
is the excess film pressure $P_F$ over the bulk pressure $P_B$ and can be measured exactly 
using a pressure cell \cite{exps}. 
\begin{equation}
P_d(l) = P_F(l) - P_B\;. 
\end{equation}
The film pressure depends explicitly on the film thickness and is given by
\begin{equation}
P_F(l) = {1\over A}{\partial\over \partial l} \ln\left(\Xi_{F}(l)\right)\;.
\end{equation}
The bulk pressure is defined in the thermodynamic limit and is given by
\begin{equation}
P_{B} = {1\over A}\lim_{l\to \infty}{\partial\over \partial l}
\ln\left(\Xi_{B}(l)\right)\;.
\end{equation}
In order to develop a systematic expansion for the 
disjoining pressure we pass to the rescaled model in terms of the rescaled field 
\begin{equation}
\phi \to {\sqrt{g}\over e\beta}\phi\;,
\end{equation}
and we measure length in terms of the Debye length $l_D$
\begin{equation}
{\bf x} \to {\bf x}\,l_D\;,
\end{equation}
where $m = \sqrt{2 \rho e^2 \beta/\epsilon}$ is the Debye mass and $l_D 
= 1/m$. The dimensionless coupling $g$ is given by
$g= l_B/l_D$ where $l_B = e^2 \beta/4\pi \epsilon$ is the Bjerrum 
length. In units of the Debye length we denote the thickness of 
the film by $L = l/l_D$. 

The disjoining pressure $P_d$ is obtained from the disjoining
pressure of the rescaled model ${\cal P}_d$ using 
\begin{equation}
P_d = m^3 {\cal P}_d = 8 \pi \rho g {\cal P}_d\;. 
\end{equation}
The effective action for the film in terms of the rescaled fields and lengths is
\begin{equation}
{\cal S}_F = -{1\over 8\pi}\int_{F_I} d{\bf x}\ \left(\nabla\phi\right)^2
+ {Z(g)\over 4\pi g} \int_{F_I} d{\bf x} \ \cos(\sqrt{g} \phi) 
-{n\over 8\pi}\int_{F_E} d{\bf x}\ \left(\nabla\phi\right)^2\;, \label{S_F}
\end{equation}
where $n = \epsilon_0/\epsilon$ and the renormalization factor $Z(g)$ simply becomes
\begin{equation}
Z(g) = {1\over \langle \cos(\sqrt{g}\phi)\rangle_B} \label{eqzg}\;,
\end{equation}
where again the above expectation is for an infinite bulk system.
The action ${\cal S}_F$ can be decomposed as 
\begin{equation}
{\cal S}_F =  {Z(g)\over 4\pi g}V_F + {\cal S}_F^{(0)} + \Delta {\cal S}_F\;,
\end{equation}
where $V_F$ is the volume of the film and the first term is the ideal 
contribution. The term ${\cal S}_F^{(0)}$ is the action for a free or 
Gaussian field theory and is given by
\begin{equation}
{\cal S}_F^{(0)} =  -{1\over 8\pi}\int_{F_I} d{\bf x}\ \left[
\left(\nabla\phi\right)^2 + \phi^2\right]  
-{n\over 8\pi}\int_{F_E} d{\bf x}\ \left(\nabla\phi\right)^2\;.
\end{equation}
The interacting part of ${\cal S}_F$ is expressed as a perturbation
\begin{equation}
\Delta {\cal S}_F = {1\over 4\pi g}\int_{F_I} d{\bf x}\ \left[
Z(g)\left(\cos(\sqrt{g}\phi) -1\right) + {g \phi^2\over 2}\right]\;,
\end{equation}
and the action ${\cal S}_B$ for the equivalent bulk system is given by
\begin{equation}
{\cal S}_B = -{1\over 8\pi}\int_B d{\bf x}\ \left(\nabla\phi\right)^2
+ {Z(g)\over 4\pi g} \int_B \ d{\bf x}\:\cos(\sqrt{g} \phi)\;. 
\end{equation}
Clearly ${\cal S}_B$ is invariant under $\phi  \to -\phi$ and hence from 
Eq. (\ref{eqzg}) one must have the expansion 
\begin{equation}
Z(g) = 1 + z_1 g + z_2 g^2 + z_3 g^3 +\ldots\;.
\end{equation}
Using the same decomposition for the bulk action as for the film action we 
obtain
\begin{equation}
{\cal S}_B = {Z(g)\over 4\pi g}V_B 
-{1\over 8\pi}\int_B d{\bf x}\ \left[
\left(\nabla\phi\right)^2 + \phi^2\right]  + 
{1\over 4\pi g}\int_B d{\bf x}\ \left[
Z(g)\left(\cos(\sqrt{g}\phi) -1\right) + {g \phi^2\over 2}\right]\;,
\end{equation}
where $V_B$ is the bulk volume. To order $g$ the action ${\cal S}_B$ may thus be written as
\begin{equation}
{\cal S}_B = {1 + z_1 g + z_2 g^2\over 4\pi g}V_B 
-{1\over 8\pi}\int_B d{\bf x}\ \left[
\left(\nabla\phi\right)^2 + \phi^2\right]  + 
{g\over 4\pi }\int_B d{\bf x}\ \left[
{\phi^4\over 4!} - {z_1 \phi^2 \over 2}\right] + O(g^2)\;.
\label{eqexpand}
\end{equation}
Hence in order to calculate $\ln(\Xi_B)$ to order $g$ one needs to evaluate
$Z(g)$ to order $g^2$. We define 
\begin{equation}
Z_n(g) = \frac{1}{\langle \cos(\sqrt{g}\beta\phi)\rangle_{n-1}}\;,
\end{equation}
where the notation $\langle O \rangle_n$ signifies the expectation value of $O$ evaluated to $n-$th
order in the cumulant expansion. Then $Z_n$ is an approximation to $Z$ correct up to and including $O(g^n)$.
Then we can write ${\cal S}_B$ correct to order $g$ as
\begin{equation}
{\cal S}_B = {Z_2(g)\over 4\pi g}V_B 
-{1\over 8\pi}\int_B d{\bf x}\ \left[
\left(\nabla\phi\right)^2 + \phi^2\right]  + 
{1\over 4\pi g}\int_B d{\bf x}\ \left[
Z_1(g)\left(\cos(\sqrt{g}\phi) -1\right) + {g \phi^2\over 2}\right]\;. \label{xib_2loops}
\end{equation} 
Clearly, calculating with the above action gives the value of $\ln(\Xi_B)$
correct to order $g$ which includes the two-loop term in the loop 
expansion coming from the term $g\phi^4/4!$ in the interaction term. We might go 
further and expand the interaction term and keep only terms of order $g$. However, when
we consider the computation of the partition function $\Xi_F$ for the film we shall see that
the naive expansion of the interaction term in this manner is illegal because it gives
rise to a spurious divergence due to the image charge singularities. In the expansion for the 
disjoining pressure we must first identify the correct Boltzmann factor associated with the image charge 
potential, which is not expansible in $g$, and only then may we expand the remaining contributions to 
order $g$ to obtain the two-loop divergence-free result for the disjoining pressure. We will find a term 
that behaves like $g\ln g$ which indicates that some contributions are indeed not expansible in $g$. 
Therefore, at two-loop order we use the form for $S_B$ given in Eq. (\ref{xib_2loops}) and a similar form 
form for $S_F$ which we discuss below. Then to $O(g)$ we have
\begin{equation}
\ln(\Xi_B) =  {Z_2(g)\over 4\pi g}V_B + \ln(\Xi_{B,0}) + \langle \Delta 
{\cal S} \rangle_{B,0} + O(g^2)\;,
\end{equation}
where 
\begin{equation}
\Xi_{B,0} = \int d[\phi] \exp\left({\cal S}_{B,0}\right)\;,
\end{equation}
with 
\begin{equation}
{\cal S}_{B,0} = -{1\over 8\pi}\int_B d{\bf x}\ \left[
\left(\nabla\phi\right)^2 + \phi^2\right]\;, 
\end{equation}
and 
\begin{equation}
\Delta {\cal S} = {1\over 4\pi g}\int_B d{\bf x}\ \left[
Z_1(g)\left(\cos(\sqrt{g}\phi) -1\right) + {g \phi^2\over 2}\right]\;.
\end{equation} 
The last term is just the first term in the cumulant expansion
about the free field theory with action ${\cal S}_{B,0}$ and is given by
\begin{equation}
\langle \Delta {\cal S}\rangle_{B,0}  = {
\int d[\phi] \exp\left({\cal S}_{B,0}\right) \Delta {\cal S}
\over \int d[\phi] \exp\left({\cal S}_{B,0}\right)}\;.
\end{equation}
Inside the bulk we define
\begin{equation}
\langle \phi({\bf x}) \phi({\bf x})\rangle_{B,0} = G_B(0)\;,
\end{equation}
by translational invariance within the bulk. This now yields
\begin{equation}
Z_1(g) = {1\over \langle \cos(\sqrt{g}\phi)\rangle_{B,0}}
=\exp\left( {g\over 2} G_B(0)\right)\;,  
\end{equation}
and 
\begin{equation}
\ln(\Xi_B) =  {Z_2(g)\over 4\pi g}V_B + \ln(\Xi_{B,0}) + {V_B \over 4\pi g} 
\left[1 +{g\over 2} G_B(0) -\exp\left({g\over 2} G_B(0)\right)\right]+O(g^2)\;.
\label{xib_result}
\end{equation}
One now carries out exactly the same calculation in the film where
within a film of thickness $L$ we define
\begin{equation}
\langle \phi({\bf x}) \phi({\bf x})\rangle_{F,0} = G_F(z,L)\;,
\end{equation}
where the field correlator at coinciding points now depends on the distance
$z$ from the surface of the film and $L$, its thickness; 
$\langle \phi({\bf x}) \phi({\bf y})\rangle_{F,0}$ is the propagator for the action ${\cal S}_{F,0}$. 
We then obtain to $O(g)$
\begin{equation}
\ln(\Xi_F) =  {Z_2(g)\over 4\pi g}V_F + \ln(\Xi_{F,0}) + 
{1 \over 4\pi g}\int_{V_F} d{\bf x} 
\left[ \exp\left(- {g\over 2} (G_F(z,L)-G_B(0))\right)
+ {g\over 2}G_F(z,L) - \exp\left({g\over 2} G_B(0)\right)\right] + O(g^2)\;.
\label{xif_result}
\end{equation} 
In the limits $L \to \infty,~z \to \infty,~z/L \to 0$, we recover the expression for $\ln(\Xi_B)$
in Eq. (\ref{xib_result}). We note that the exponent of the first exponential under the integral
sign in Eq. (\ref{xif_result}) contains the repulsive image charge potential which is singular as $z \to 0$.
Evidently, it is illegal to expand this exponential as it stands and it is clear that it contains the
correct Boltzmann factor for the potential due to the image charges. In terms of the rescaled system the 
disjoining pressure is then given by
\begin{equation}
\beta {\cal P}_d  = {\partial \over \partial L} B_0(L) +
{\partial \over \partial L} B_1(L)\;, 
\end{equation}
where 
\begin{equation}
B_0(L) = {1\over A}\left(\ln\left(\Xi_{F,0}(L)\right) - L\lim_{L'\to \infty}
{\ln\left(\Xi_{B,0}(L')\right)\over L'}\right)\;,
\end{equation}
is the one-loop or $O(1)$ contribution in the expansion in $g$, and $A$
is the area of the film. The two-loop term, or term to $O(g)$, is given by
\begin{equation}
B_1(L) = {1 \over 4\pi g}\int_0^L dz
\left[ \exp\left(- {g\over 2}G_R(z,L)\right) - 1 + {g\over 2}G_R(z,L)\right]\;,
\label{eqb1}
\end{equation}
with 
\begin{equation}
G_R(z,L) = G_F(z,L)- G_B(0)\;.
\end{equation}

We note that there is no need to calculate $Z_2(g)$, the approximation to $Z(g)$ accurate to 
$O(g^2)$, in order to compute the disjoining pressure, since this term is the same in both
the film and the bulk and hence cancels identically. We find 

\begin{equation}
G_R(z,L) = \int_1^\infty dP \left[ {2 \Gamma^2(P) \exp(-2PL) 
+ \Gamma(P) \exp(-2P(L-z)) + \Gamma(P) \exp(-2Pz)
\over 1 - \Gamma^2(P) \exp(-2PL)}\right]\;, \label{eqgr}
\end{equation}
where 
\begin{equation}
\Gamma(P) = {P - n \sqrt{P^2-1}\over P + n \sqrt{P^2-1}}\;.
\end{equation}
We can compare the result in Eq. (\ref{eqb1}) with the result of Netz \cite{netz2} where
he computes the two-loop contribution to the free energy of the film. When 
normalized with respect to the bulk free energy, the result of Netz agrees with
Eq. (\ref{eqb1}) expanded to and truncated at $O(g)$. In the case where $\Delta=0$ the expression of Netz is indeed 
the disjoining pressure correct to two-loops. However, in the case where $\Delta = (1-n)/(1+n) > 0$, 
{\em i.e.} when there is dielectric discontinuity in the system, the expression truncated at $O(g)$ is divergent.  
As already remarked, this is due to the presence of image charges which repel charge away from the surface 
into the film when $\Delta > 0$ as is the case of air/water films, for example. Netz obtains a finite result
using this truncated expansion by introducing an artificial Ultra-Violet cut-off in the $P$ integration
defining $G_R(z,L)$ in Eq. (\ref{eqgr}). The UV cut-off is associated with a microscopic length scale such as
the ionic radius of the solute. Expanding the exponential in the expression for $B_1$ amounts to an 
illegal expansion of a Boltzmann weight which keeps charge away from the surface. This term needs to be 
treated with care as in the Onsager-Samaras \cite{onsa} calculation of the excess surface tension of ionic
solution where a similar term also arises.  

We write
\begin{equation}
\beta {\cal P}_d = \beta {\cal P}^{(0)}_d + \beta {\cal P}^{(1)}_d\;,
\end{equation}
where ${\cal P}^{(n)}_d$ is the $O(g^n)$ contribution to ${\cal P}_d$.
The one-loop, $O(1)$, contribution to the disjoining pressure has been worked
out by many authors and is given by
\begin{equation}
\beta {\cal P}^{(0)}_d = -{1\over 2\pi} \ \int_1^\infty P^2 dP
{\Gamma^2(P) \exp(-2PL)\over 1 - \Gamma^2(P) \exp(-2PL)}\;.
\label{oneloop}
\end{equation}
From above we find the two-loop contribution to be given by
\begin{equation}
\beta {\cal P}^{(1)}_d = {1\over 4\pi g}\left( \exp\left( -{g\over 2} G_R(L/2,L)\right)  
+ {g\over 2} G_R(L/2,L) -1\right) + {1\over 4 \pi} \int_0^{L/2}
dz {\partial \over \partial L}G_R(z,L) \left[1 - \exp\left( -{g\over 2} G_R(z,L)\right)\right]\;.
\label{eqp1}
\end{equation}
We emphasize that this expression is finite for all $L > 0$ and requires no UV regularization.

We may attempt to expand Eq. (\ref{eqp1}) to $O(g)$ to get  
\begin{equation}
\beta {\cal P}^{(1)}_d = {g\over 32 \pi} G^2(L/2,L) + {g\over 8\pi}\int_0^{L/2}
dz G_R(z,L){\partial \over \partial L}G_R(z,L) + O(g^2)\;, \label{eqp12}
\end{equation}
thus giving what naively looks like the two-loop contribution. 

The function $G(z,L)$ can be shown to be finite every where for $z\in [0,L/2]$ except at $z=0$. 
It is straightforward to isolate the divergent part of $G_R(z,L)$ near $z=0$, and we find that
\begin{equation}
G_R(z,L) = \Delta {\exp(-2z)\over 2z} +  G'_R(z,L)\;,\label{eqgrp}
\end{equation}
where $G'_R(z,L)$ is finite for all $z\in [0,L/2]$. Hence, the first term of 
Eq. (\ref{eqp12}) is finite but the second has a logarithmic divergence
at $z=0$ when $\Delta > 0$. Introducing a large $P$ cut-off in the integral in Eq. (\ref{eqgr})
defining $G_R(z,L)$, as was done in \cite{netz2}, eliminates the $1/z$ singularity in $G(L,z)$ 
but has no physical basis, in fact it is eliminating a physical effect which should be there due 
to the presence of image charges; the divergent term of $G_R(z,L)$ should be kept in
the exponential (which is just the Boltzmann factor due to the image charges).
We can now carry out a legal expansion in $g$ for the terms which give finite contributions, 
and obtain
\begin{equation}
\beta {\cal P}^{(1)}_d = {g\over 32 \pi} G_R^2(L/2,L) + {g\over 8\pi}\int_0^{L/2}
dz\  G'_R(z,L){\partial \over \partial L}G'_R(z,L) + 
+ {1\over 4\pi}\int_0^{L/2} dz\  {\partial \over \partial L}G'_R(z,L)
\left[ 1- \exp\left(-{g\Delta e^{-2z}\over 4 z}\right)\right]\;.
\label{eqpint}
\end{equation}
We now write the two-loop contribution to the disjoining pressure as the
sum of these three terms
\begin{equation}
\beta {\cal P}^{(1)}_d = \beta {\cal P}^{(1a)}_d + \beta {\cal P}^{(1b)}_d
+ \beta {\cal P}^{(1c)}_d\;. \label{three_terms}
\end{equation}
The first two terms can be immediately written down and a careful asymptotic
analysis of the last term of Eq. (\ref{eqpint}) gives a finite resummed 
two-loop contribution
\begin{eqnarray}
\beta {\cal P}^{(1a)}_d &=& {g\over 32 \pi} G_R^2(L/2,L)\;, \label{p1a}\\
\beta {\cal P}^{(1b)}_d &=& {g\over 8\pi}\left[\int_0^{L/2}
dz\  \left( G_R(z,L)({\partial \over \partial L}G'_R(z,L)-
{\partial \over \partial L}G'_R(0,L)) +
G'_R(z,L){\partial \over \partial L}G'_R(0,L)\right)\right]\;, \label{p1b}\\
\beta {\cal P}^{(1c)}_d &=& {g \Delta \over 16 \pi}\left[ {\partial \over \partial L}G'_R(0,L)
\left( - \ln({g\Delta\over 2}) + 1 - 2 \gamma\right)\right]\;,\label{p1c}
\end{eqnarray}
where $\gamma = 0.577215....$ is Euler's constant. We see from the third term a contribution
proportional to $g\ln g$ signifying that the assumption that ${\cal P}_d$ can be naively expanded
in $g$ is incorrect.

\section{Large $L$ asymptotics}
In this section we will analyse the asymptotic behavior of the 
disjoining pressure of thick films. We start by recalling the one loop
result. At large $L$ the behavior of the integral determining
${\cal P}^{(0)}_d$ in Eq. (\ref{oneloop}) is dominated by small $P$,{\em i.e.}
the contribution to the integration coming from near $P=1$. The leading term for large $L$ is
\cite{mani}
\begin{equation}
\beta {\cal P}^{(0)}_d \approx - {\exp(-2L)\over 4 \pi L}\;.
\end{equation}
We see that the thick film disjoining pressure at one-loop is attractive and
exponentially screened with characteristic length $l_D$. Interestingly,
the value of $\Delta$ does not appear in the leading term of this 
expression which is therefore independent of the strength of the 
dielectric discontinuity between the film and exterior. For large $L$ we find that
\begin{equation} 
G_R(L/2,L)\approx 2{\exp(-L)\over L}\;,
\end{equation}
and thus 
\begin{equation}
\beta {\cal P}^{(1a)}_d \approx {g\over 8 \pi} {\exp(-2L)\over L^2}\;,
\end{equation}
which is clearly negligible with respect to the one-loop contribution at large $L$.

The behavior of $G'(z,L)$ at large $L$ is given to leading order
by
\begin{equation}
G'_R(z,L) \approx \exp(-2L)\left[ {1\over L} + {1\over2}{1\over L-z}\exp(2z) + 
{1\over2}{1\over L+z}\exp(-2z)\right] + H(z)\;,
\end{equation}
where
\begin{equation}
H(z) = \int_1^\infty dP\; (\Gamma(P)-\Delta)\exp(-2Pz)\;.
\end{equation}
Hence
\begin{equation}
{\partial \over \partial L} G'_R(z,L) \approx -2\exp(-2L)
\left[ {1\over L} + {1\over2}{1\over L-z}\exp(2z) + {1\over2}{1\over L+z}\exp(-2z)\right]\;.
\end{equation}   
Using this result we find that for large $L$ the leading behavior is
\begin{equation}
\beta {\cal P}^{(1b)}_d \approx -{g \over 16 \pi}\ {\exp(-2L)\over L}
\left[\ln(L) + \gamma + {8\over 1-n^2}\left( \ln(1+n) - n \ln(2)\right) + 
O\left(\frac{1}{L^{1/2}},\frac{\ln(L)}{L}\right)\right]\;,\label{p1b_asy}
\end{equation}
and for the special case $n=0$ the non-leading $O(1/L^{1/2})$ correction is absent and
this result becomes
\begin{equation}
\beta {\cal P}^{(1b)}_d \approx -{g \over 16 \pi}\ {\exp(-2L)\over L}
\left[\ln(L) + \gamma + \frac{1}{2L}\ln(L) +  O\left(\frac{1}{L}\right)\right]\;. \label{p1b_n=0}
\end{equation}
It can be shown that the coefficient of $\exp(-2L)\ln(L)/L$ is independent of $n$ in the general case and
is therefore given for any value of $n$ by this result. Hence, as in the case of ${\cal P}^{(0)}_d$, the
leading contribution to ${\cal P}^{(1)}_d$ at large $L$ is also independent of $\Delta$.

We also find 
\begin{equation}
\beta {\cal P}^{(1c)}_d \approx -{g \Delta\over 4 \pi}\ {\exp(-2L)\over L}\
\left[- \ln(g\Delta/2) + 1 - 2\gamma) \right]\;. \label{p1c_asy}
\end{equation}
We notice that if we take the contribution of the one-loop term with the leading 
two-loop term then we have
\begin{equation}
\beta {\cal P} \approx - {\exp(-2L)\over 4 \pi L}\left[ 1 + {g\over 4} \ln(L)\right]\;,
\end{equation}
which suggests the speculative resummation
\begin{equation}
\beta {\cal P} \approx - {\exp(-2L)\over 4 \pi L^{1- {g/4}}}\;.\label{exponent}
\end{equation}
If valid, the terms leading to this resummation must come from higher-order terms in the
cumulant expansion. It can be verified that they do not arise from higher-order terms in $g$ in
the expansion of Eq. (\ref{eqp1}).

Examining the contributions to $\beta{\cal P}^{(1)}$ as defined by
Eq. (\ref{three_terms}) at large $L$ we see that  $\beta{\cal P}^{(1a)}$ is clearly
subdominant, but although $\beta{\cal P}^{(1b)}$ is clearly the
leading term at very large $L$ we must bear in mind that this 
expansion is valid for small $g$. Comparing 
Eq. (56) with Eq. (57) we see that there exists a cross-over length 
$L^*$ defined by
\begin{equation}
L^* \sim 1/g^4, \label{eqlstar}
\end{equation}
such that for $1 \ll L^* \ll L$ we have
$\beta{\cal P}^{(1)} \approx \beta{\cal P}^{(1c)}$ and thus in this region
$\beta{\cal P}^{(1)}$ has the same functional form a the one-loop 
result. Thus only for $L \gg L^*$ does one see the modification of the 
functional form of $\beta{\cal P}$ with respect to the one-loop result.  
 
\section{Numerical Results}
The major result of this paper for the large-$L$ asymptotic behaviour of ${\cal P}_d$ is the 
expression Eq. (\ref{p1b_asy}). In Fig. (\ref{R1b_L_fig}) we show data points for 
\begin{equation}
R^{(1b)}_d(L)~=~\frac{16\pi L}{g}e^{2L}\beta{\cal P}_d^{(1b)} \label{R1b_L}  
\end{equation}
versus $L$ computed numerically using Eqs. (\ref{eqgr}), (\ref{eqgrp}) and (\ref{p1b}) for $n=0$. The solid line
is the asymptotic form derived for $R^{(1b)}_d(L)$ from the expression in Eq. (\ref{p1b_n=0}) but including a 
term proportional to $1/L$ whose coefficient we determine by a fit to the data. This fit curve is
\begin{equation}
R^{(1b)}_d(L)~=~-\gamma - \ln(L) - \frac{1}{2L}\ln(L) - \frac{1.45354}{L}\;. \label{R1b_L_n=0}
\end{equation}

It is not useful to directly plot ${\cal P}^{(1)}_d$ given in Eq. (\ref{eqp1}) since it is a rapidly changing function of $L$.
However, the important comparison is with the one-loop result ${\cal P}^{(0)}_d$, which corresponds
to the attractive Casimir force and which is present even as $\rho \to 0$, that is, $g \to 0$. 
In this limit ${\cal P}^{(0)}_d$ arises solely from the thermal fluctuations of the electrostatic field in the presence of 
interfaces of dielectric discontinuity. In the limit $g \to 0$ the two-loop contribution must vanish since the one-loop
result is exact; this can be seen in the expressions given above. However, for $g > 0$ we see that
${\cal P}^{(1)}_d$ is of the same sign as ${\cal P}^{(0)}_d$ and so corresponds to an increase in the attractive
force. In Fig. (\ref{one_two_fig}) we plot the ratio ${\cal P}^{(1)}_d/{\cal P}^{(0)}_d$ for various values
of $g$: $g=0.1,0.3,0.5$. We see that ${\cal P}_d^{(1)}$ is comparable with ${\cal P}_d^{(0)}$ for $g \sim 0.5$ 
over a wide range of values of the film thickness $L$. For small enough $L$ ($L \le 2$) the one-loop result 
eventually dominates.  It is clear that ${\cal P}_d^{(1)}$ scales, as expected, approximately linearly with $g$ in the
small range considered. The value $g \sim 0.5$ corresponds to a solute density in water of about 50 mM at room temperature.
This result shows that a quantitative analysis of the collapse phenomenon in a thin electrolyte soap film
must take account of higher-order effects for solute densities greater than 10 mM. An important point to note
is that both ${\cal P}^{(0)}_d$ and ${\cal P}^{(1)}_d$ are negative for all $L>0$, and so the two-loop contribution adds
to the attractive force between the interfaces.

In Fig. (\ref{ratio_fig}) we plot the ratio  ${\cal P}^{(1)}_d/{\cal P}^{(1c)}_d$ where the denominator
is the approximation to ${\cal P}^{(1)}_d$ given in Eq. (\ref{p1c_asy}). This approximation arises from
the resummed terms which properly account for the Boltzmann factor associated with the image charges. 
It is clear that this approximation is a good one over a wide range of film thickness $L$ and shows
that the dominant contribution in this range behaves as $g\ln(g)$. It is clear that this should be the case 
in this range as in the case $g = 0.5$, from Eq. (\ref{ratio_fig}) $L^* \sim 16$.

\section{Conclusion}
In this paper we have carefully calculated the two-loop contribution, ${\cal P}^{(1)}_d$, to the
disjoining pressure for a thin electrolyte soap film of thickness $l$ and monovalent solute density $\rho$.
The surfaces of the film are the interfaces of dielectric discontinuity separating the film interior of
dielectric constant $\epsilon$ from the exterior which consists of non-ionic medium of dielectric constant 
$\epsilon_0$. The length scales in the problem are set by the Bjerrum length $l_B = e^2\beta/(4\pi\epsilon)$
and the Debye length $l_D^{-1} = m = \sqrt{2\rho e^2\beta/\epsilon}$. The model used for the ionic interactions
is the Sine-Gordon field theory extensively discussed in earlier work \cite{deho}. All quantities can be expressed
in dimensionless form in terms of the coupling constant $g = l_B/l_D$ and $n=\epsilon_0/\epsilon$, with overall 
dimensions carried by the Debye mass $m$. By naive power counting the $N-$loop contribution is the $O(g^{N-1})$. 
The self-energy divergences are included in the explicit renormalization constant $Z(g)$ defined in Eq. (\ref{zed}).

We have computed ${\cal P}^{(1)}_d$ in the cumulant expansion of the film free energy $\ln(\Xi_F)$ given in terms
of the effective action $S_F(\phi,L)$ defined in Eq. (\ref{S_F}). Although, we are naively working to $O(g)$ it
is clear that it is illegal to expand all terms to this order since the effect of image charges is to 
introduce a spurious Ultra-Violet divergence in this case. As an ion at distance $z$ from an interface approaches
the interface the image potential is repulsive and diverges like $\beta V_I(z) \approx g\Delta/2z$, where $\Delta = (1-n)/(1+n)$. 
It is necessary to keep the Boltzmann factor $\exp(-\beta V_I(z))$ intact for this otherwise divergent contribution,
which is then regulated in the proper way. The resulting expression for ${\cal P}^{(1)}_d$ is given in Eq. (\ref{eqpint}).
The outcome is that the naive expectation that this contribution is $O(g)$ is incorrect and the asymptotic analysis of
the expression for ${\cal P}^{(1)}_d$ gives the term ${\cal P}^{(1c)}_d$ shown in Eq. (\ref{p1c}) which contains
a term of order $g\ln(g)$. Moreover, in Fig. (\ref{ratio_fig}) we see that the approximate expression  ${\cal P}^{(1c)}_d$
contributes the major part of the full result for small $L = l/l_D$, and so the collapse phenomenon observed in
thin electrolytic soap films may be modified significantly by such non-analytic terms. From Fig. (\ref{one_two_fig})
we observe that the two-loop and one-loop contributions are comparable for $g \sim 0.5$, corresponding to solute 
densities $\rho \sim 50 mM$. For $L \ll 1$ we would still expect the one-loop contribution, ${\cal P}^{(0)}_d$, to dominate 
all others on physical grounds; the image charges will ensure that all ions are expelled from the film leaving 
a non-ionic water-filled film. The only, and therefore exact, contribution is the Casimir term due purely to
the presence of the dielectric discontinuities.

The question arises how large $g$ can be taken before the loop expansion breaks down. The only evidence is from
the calculation of the bulk Debye pressure where the Debye-H\"uckel correction to the ideal value $P_0$ is
\begin{equation}
P_B~=~P_0(1-\frac{g}{6})\;.
\end{equation}
This suggests that the expansion parameter might be as small as $g/6$. However, we know that the two-loop term is
$O(g\ln(g))$ and that our calculation deals with the effects of an interface whose physics is
completely different from that of the bulk. Nevertheless, we may take this result as an indicative of what we might
expect.

We might have expected that the leading large-$L$ contribution from ${\cal P}^{(1)}_d$ behaves in a similar manner 
to that of the one-loop contribution, namely,
\begin{equation}
{\cal P}^{(0)}_d~\approx~-\frac{A}{L}\exp(-2L)\;.
\end{equation}
However, the analysis of the large $L$ behaviour shows a departure from expectations because of subtle contribution
from the image charges; we see from Eq. (\ref{p1b_n=0}) and from Fig (\ref{R1b_L_fig}) that the leading large-$L$ 
term behaves as
\begin{equation}
{\cal P}^{(1)}_d~\approx~-gA^\prime\frac{\ln(L)}{L}\exp(-2L)\;.
\end{equation}
This result implies that for sufficiently large $L$ the loop-expansion fails to converge since ${\cal P}^{(1)}_d$
will become larger than ${\cal P}^{(0)}_d$. We suggest that higher-order loop contribution can be resummed to give
an anomalous exponent as shown in Eq. (\ref{exponent}); this point requires further investigation. However, for
physically relevant conditions the loop expansion will be valid.

The conclusion of this paper is that the naive loop expansion of the free energy for these systems is invalid in
the presence of a dielectric discontinuity, and a careful resummation of higher loop-order terms must be employed to 
properly take into account the Boltzmann weight of image charges. Furthermore, at two-loop order as generated by the
cumulant expansion, the disjoining pressure is finite without the need for the introduction of an Ultra-Violet 
regulator. In addition, we have shown that the two-loop disjoining pressure is negative for $L > 0$ and so
enhances the attraction between the interfaces predicted at one-loop order.

\pagestyle{plain}

\end{document}